# Electro optic effect measurements for waveguide inscribed in X-cut LiNbO$_3$ substrate using femtosecond laser direct writing process

Amar Ghar*, Sanyogita, Utpal Das and P. K. Panigrahi

Femtosecond Laser Fabrication Lab, Centre for Lasers and Photonics, Indian Institute of Technology Kanpur, Uttar Pradesh- 208016 India.
amarghar@iitk.ac.in

**Abstract:**
This work represents a particular application of waveguide fabricated by femtosecond laser micromachining technology. More specifically, we report the development of an optical modulator based on the fabrication of single mode optical waveguide structures buried in X-cut lithium niobate crystal with the femtosecond laser direct writing method. Here, change in refractive index profile is measured using near field intensity profile measurement method at optimized writing conditions. It has been observed that the change refractive index ($\Delta n$) is in the range of $10^{-4}$. Finally, the behavior of femtosecond written waveguides as an intensity modulator at 632.8 nm and 1550 nm under the influence of an external electric field is analyzed by pattering electrode structures on the substrate respectively.

**Introduction:**
LITHIUM NIOBATE (LiNbO$_3$) is a key material for active optical waveguide applications used in photonics because of its high electro-optic, acousto-optic, and optical nonlinear coefficients [1]. Titanium or zinc ion in diffusion [2], proton exchange [3], and ion-implantation [4] are the traditional techniques most widely used to fabricate optical waveguides in LiNbO$_3$, which is a basic structure needed for optical integrated circuits (OICs). Femtosecond (fs) laser direct writing on a transparent material is being used for waveguide fabrication for different optical devices such as power splitters, interferometers, waveguide lasers, and optical sensors [5, 6]. The advent of femtosecond laser micromachining for fabrication on transparent material has opened up new avenues for the development of optical waveguide-based photonic components [7]. Compared to traditional fabrication techniques, fs-laser waveguide writing has a number of advantages – simple, cost effective, and capable of a wide variety of material processing. Fs-laser micromachining can be used for the fabrication of both microchannels and waveguides [5-7].

To date, direct writing using femtosecond laser has been demonstrated in order to fabricate optical waveguides on different transparent materials like LiNbO3, glasses, polymers, active material, etc. [7, 8]. Specifically, fabrication of waveguide based microstructures in nonlinear crystals like LiNbO$_3$, KDP, KD*P, etc has been attracted great attention because of the promising potential application as optical coupler [9], beam splitter [10], fork grating [11], on-chip nonlinear optical processing, quantum information processing, etc. Femtosecond laser based waveguide fabrication provides feasibility towards development of both double-line waveguides [12] and depressed cladding waveguides [13] inside transparent materials including LiNbO$_3$ crystals respectively. The optical waveguide structure inside the LiNbO3 crystal has been developed by the local modification of the refractive index by following a variety of mechanisms at tightly focused region of femtosecond pulses inside the material. The combination of nonlinear absorption through photo-ionization and avalanche ionization allows energy to be deposited in a small volume around the focus [13, 14]. The exact physical mechanism behind refractive index changes in substrate materials during femtosecond laser interaction is not yet fully understood and is still being investigated. LiNbO$_3$ materials are chemically inert and transparent for a wide range of wavelengths therefore writing waveguides in this material is particularly attractive for the development of optical modulators.

In this work, we present optical waveguide fabrication on an X-cut LiNbO$_3$ crystal substrate using femtosecond laser based fabrication technology. The fabrication parameters such as fs-laser pulse energy, pulse duration, repetition rate, substrate translation speed, and focusing optics are optimized in such a way that the waveguide will support both the 633nm and 1550nm ($\lambda$) respectively. The X-cut LiNbO$_3$ has been used to facilitate the laser fabrication of the waveguide through patterned electrodes for electro-optic devices. Finally, the electro-optic properties including half-wave voltage and optical losses at 632.8nm and 1550nm have been measured for the developed prototype.

**Device Fabrication:**

The electrode structure required for this electro-optic modulation has been deposited on the surface of an X-cut LiNbO3 crystal using photolithography followed by an electron beam deposition technique. The photolithography mask used has been a set of ten opaque 8μm wide lines separated by 250 μm. The overall photolithography process flowchart is shown in Fig. 1. A thin layer of $ZrO_2$ (50nm), Ti (50nm) and Au (200 nm) has been deposited by using an E-beam vacuum evaporator, followed by lift-off in acetone for 30 minutes.

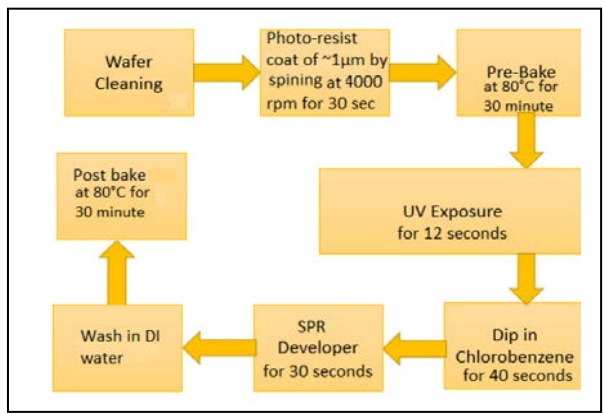

Fig. 1: Flow-chart: Photolithography Process

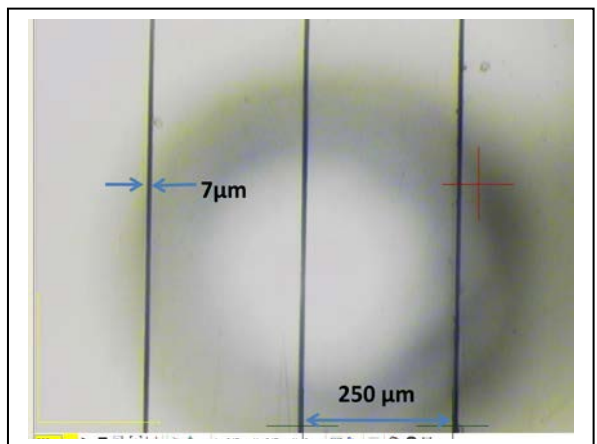

Figure 2: Final processed sample (Microscopic image)

The desired pattern achieved is shown in Fig. 2. The electrode configuration parameters have been analyzed and confirmed using an optical surface profiler as shown in Fig. 3.

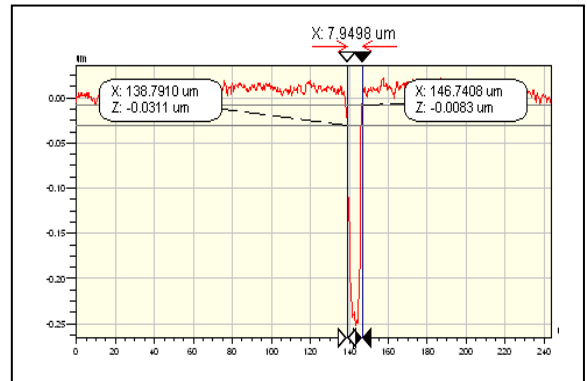

Figure 3: Surface profile including 8 μm wide spacing

Once the desired pattern on the substrate is formed, 8μm windows in the Au pattern have been exposed to 120 fs-laser pulses with a 1 kHz rep rate and 775 nm central wavelength (CLRK-MXR, CPA Series) for the fabrication of optical waveguide structure. The x-cut $LiNbO_3$ crystal used in this work has been cut into 14mm × 3mm pieces and optically end polished. The experimental setup is shown in Fig.4 below.

The $LiNbO_3$ substrate that has been placed on the X-Y motorized stage has now been exposed to the laser beam focused by a 10× lens. The waveguides fabricated by fs-laser writing had been obtained at average powers of 2 mW, 2.5 mW, and 3 mW respectively at a scan speed of 50μm/sec. The femtosecond laser machining parameters including laser power, scan speed has been optimized based on experimental trials and literature data.

**Optical Waveguide Characterization**

In order to analyze the properties of the fabricated waveguide structures, the end fire waveguide coupling measurement setup was developed. The transmitted beam near-field intensity method [15] is used in this work to estimate optical losses and refractive index change in the femtosecond exposed tracks. In this method, the transmitted beam at 632.8 nm and 1550 nm from the end facet of the optical waveguide are scanned by a CCD camera, and intensity is measured as shown in Fig. 5 below. An optical waveguide quality is assured by measuring optical losses at 632.8 nm and 1550 nm at different waveguide respectively using photodiode are summarized in table below:

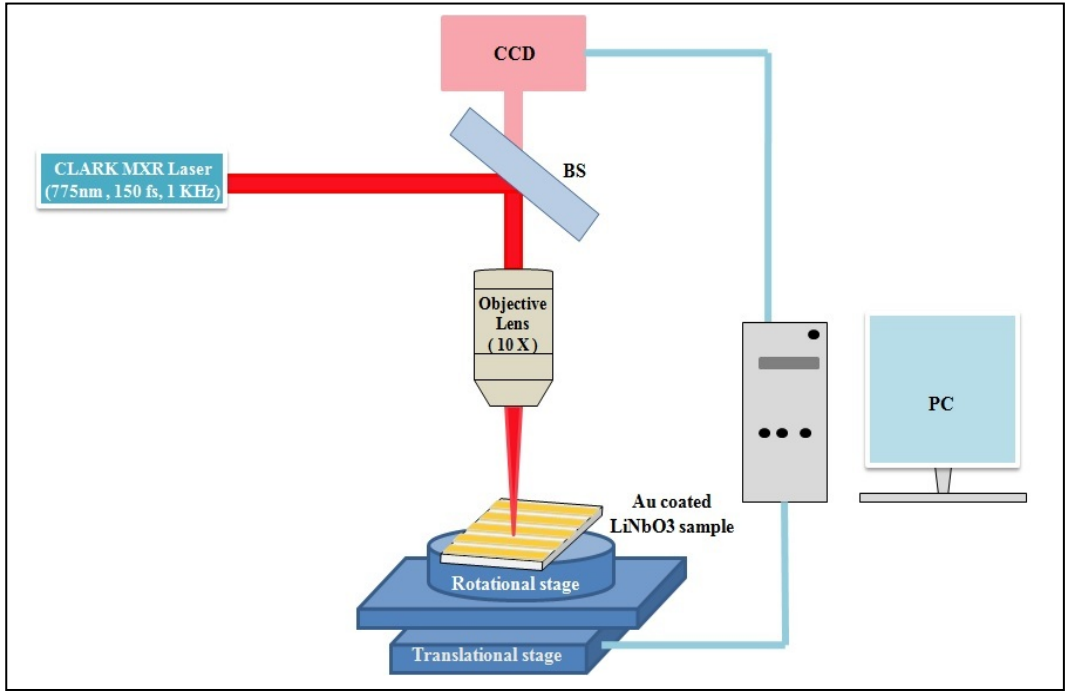

Figure 4: Experimental set up of femtosecond laser based micromachining system

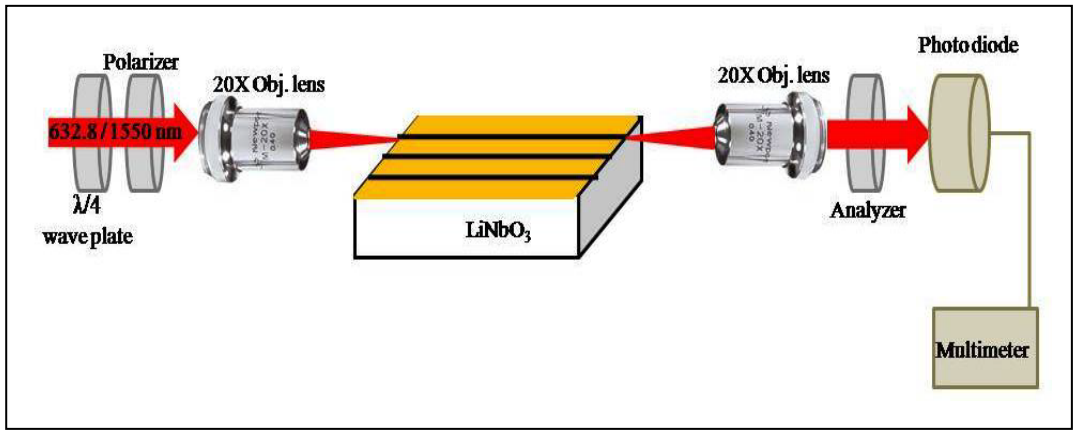

Figure 5: Experimental set up for optical characterization

| Waveguide | Optical Loss @ 632.8 nm (dB/cm) | Optical Loss @ 1550nm (dB/cm) |
|---|---|---|
| A | 1.857 | 2.29 |
| B | 2 | 1.6 |
| C | 2.07 | 1.7 |

Table1 : Optical loss measurement @632.8 nm and 1550 nm

(In the table 1 optical waveguide named as A, B and C represents waveguides fabricated at femtosecond laser power 2 mW, 2.5 mW and 3 mW respectively.)

One can detect simultaneously the laser beam with Gaussian shape in the both x and y directions which are coming out from the waveguide. Near field intensity profile is measured at 632.8 nm using CCD replacing photodiode in the experimental set up shown in Fig. 5

The transmitted beam near-field method is direct technique to deduce the index distribution from optical measurements. Here, the transmitted beam intensity is measured near the end facet and from which refractive index distribution can be calculated [16] is employed in this work to estimate refractive index change in the femtosecond exposed tracks. In this method, the intensity profile from the optically polished end facet of the optical waveguide at 632.8nm is captured by a CCD camera. Normalized electric field component A(x, y) was determined for measured intensity. In case when the refractive index change is very small, change in refractive index along the cross-section of the waveguide is calculated with help of image analysis using beam profiling software as:

$$\Delta n\,(x,y) = \sqrt{n_s^2 - f\frac{1}{K^2 A}\nabla_t^2 A} - n_s$$

Where, $n_s$ is the x- or y- directional refractive index of $LiNbO_3$, f is the filter cut-off frequency which for this work is 1 since the data is not filtered, k is the free space propagation constant, A(x,y) is the normalized field amplitude respectively.

The laser intensity and corresponding estimated refractive index profile at the output cross section of waveguide structures fabricated inside the X-cut $LiNbO_3$ sample with femtosecond laser average power 2.5 mW and 3 mw at 50 um/sec scan speed is as shown in Fig. 6 and 7 below:

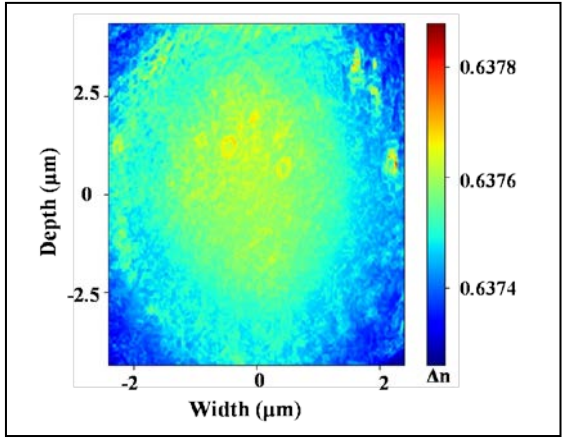

Figure 6 : Refractive index profile of waveguide at 632.8 nm

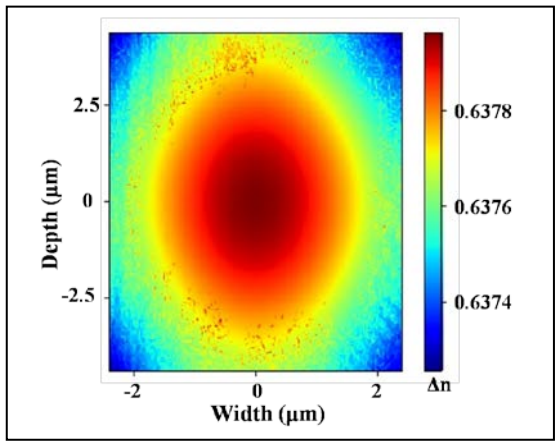

Figure 7 : Refractive index profile of waveguide at 632.8 nm

As shown in Fig. 6 and 7, the refractive index change is very small ~ $10^{-4}$ throughout the cross section of waveguide in both the cases. Also the variation of Δn is more uniform in case of waveguide fabricated at 3 mW femtosecond laser average power compare to that of developed at 2.5 mW respectively. This study also signifies the sensitivity of waveguide properties on the femtosecond laser parameters.

**Electro-optic effect measurement:**

Finally, the performance of developed device as a fully integrated tunable electro-optic modulator has been examined by applying a dc voltage to microelectrodes in a range from 0 to 20 V. To apply DC voltage across the electrode, the electrode is connected to the power supply with the help of microelectrode contact made by 10 μm thick gold wire. The electrode wire bonding is achieved by a west bond microelectrode facility. Bonding quality is assured by checking connectivity and microscopic analysis of electrode structures as shown below:

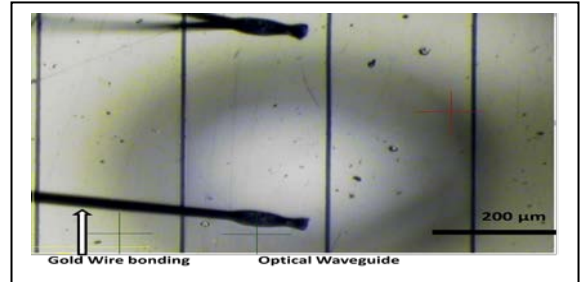

Figure 8: Electrode contacts bonding(Microscopic image)

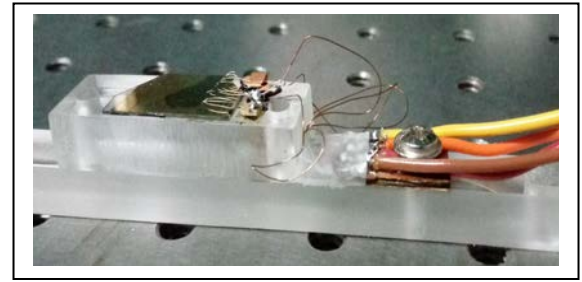

Figure 9: X-cut $LiNbO_3$ based linear optical modulator (Complete device)

The experimental setup for measuring the electro-optic coefficient is as shown in Fig. 10. The laser beam is polarized at an angle of $45^0$ with respect to the crystal axis with the help of a half-wave plate. In this configuration, the beam propagates along the optical axis. The electrodes are placed in a direction such that the electric field is applied transverse to the direction of the beam propagation, as shown in Fig. 10. In case of absence of external voltage, both the polarization components of incident light propagating through the waveguide travels at the same velocity. However, when external voltage has been applied, a phase shift is introduced between components ($E_x$ and $E_y$) of light due to birefringence and electro-optic properties of $LiNbO_3$. The beam at the output of the waveguide is analyzed by a Glan Thompson polarizer which is placed at a cross-position with respect to the input polarizer. Light from the analyzer is detected by an appropriate detector. The measured intensity after the achromatic analyzer, from the photo detector, is plotted as a function of the applied bias as shown in Fig. 11 and 12.

As shown in Fig. 11 and 12, the half wave voltage ($V_\pi$) measured at 632.8nm and 1550nm wavelengths are found to be ~3.5 V and 9 V, respectively.

**Electro-optic coefficient measurement:**
Based on experimental results obtained in case of optical waveguide based linear modulator, electro-optic coefficients for waveguide fabricated on an X-cut LiNbO$_3$ crystal can be determined as [17, 18]

LiNbO$_3$ crystal developed by femtosecond laser based micromachining technology has been demonstrated. To apply an external electric field, the electrode structure is developed by patterning

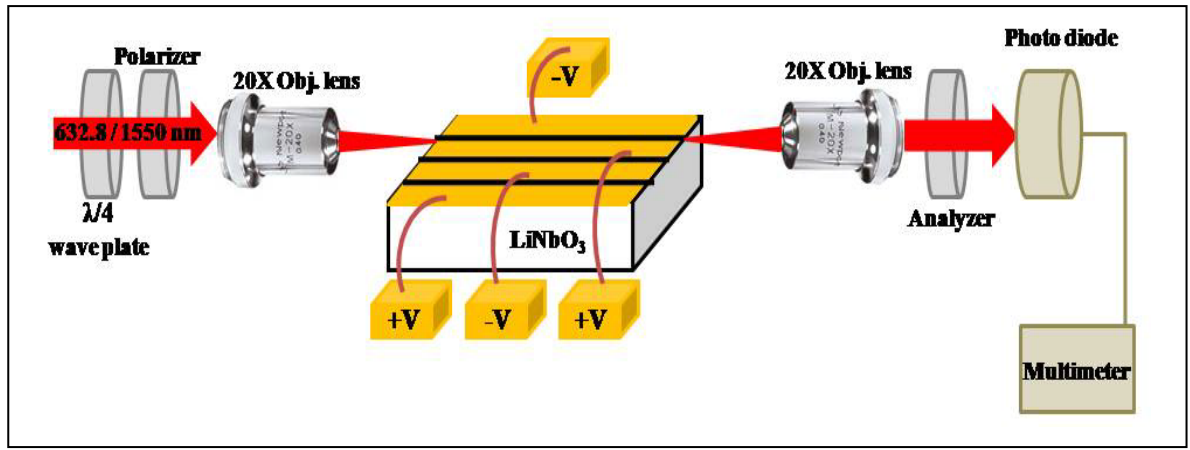

Figure 10: Experimental set up for Electro- optic effect characterization

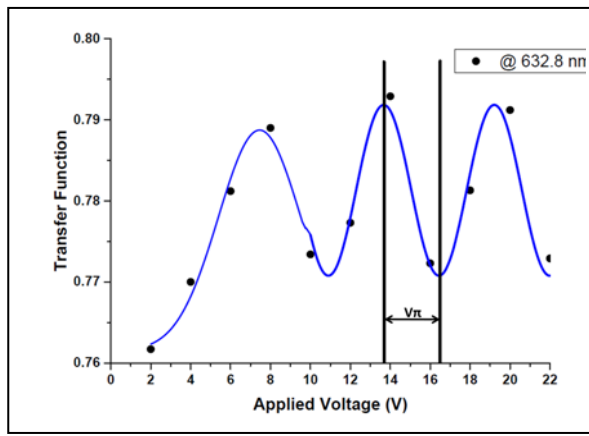

Figure 11: Electro optic effect measurement @ 632.8nm

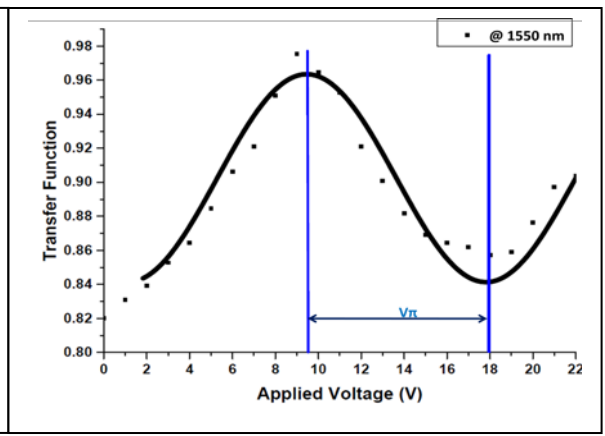

Figure 12: Electro optic effect measurement @ 1550 nm

$$V_\pi = \frac{\lambda_g \, d}{n_e^3 r_c \tau_{TE} L}$$

Where, $\lambda_g$ is 632.8 nm and 1550 nm wavelength of optical signal; d is 10 μm separation between electrode; $n_e$ is 2.21 an extra ordinary refractive index along the direction of applied field; $r_c$ is electro-optic coefficient; $\tau_{TE}$ be overlap factor of electric and optical field along the waveguide structure, in our case it is considered as 1 and L be interaction length of optical and electrical signal.

The experimental measurement of $V_\pi$ leads to value of electro–optic coefficient of waveguide comes as 16.744 pm/V for 632.8 nm and 15.955 pm/V for 1550 nm optical signal respectively.

**Conclusions:**
In this paper, a working electro-optic effect phase shifter based on optical waveguides in X-cut ZrO$_2$, Ti, and Au where Au is the electrode material, whereas, Ti is used for adhesion on ZrO$_2$ as the adhesion of Au on the LiNbO$_3$ substrate is poor. But, Ti diffusion into the LiNbO$_3$ substrate is undesirable here, hence, a passive layer of ZrO$_2$ is deposited first. An optical waveguide which allows the mode propagation at the 632.8nm and 1550nm with low optical losses has been developed by optimizing femtosecond laser micromachining parameters. The Refractive index profile of waveguide structure is measured with refractive index change in the range of $10^{-4}$ for horizontal and vertical direction for the wavelengths of 632.8 nm. Electro-optic effects and the corresponding value of coefficients are measured by applying a dc voltage on the pre-patterned Au electrodes from the response observed as the expected $(\sin)^2$ curve.

**Declaration of Competing Interest**
The authors declare that they have no known competing financial interests or personal relationships that could have appeared to influence the work reported in this paper.

**Acknowledgements**
Authors acknowledge support of Mr. Sunil Mugdum CMTI Bangalore, Mr. Dharmendra Malik for development of electrode structure and appreciate Mr. Devendra Singh for technical support.